# Exciton-Driven Renormalization of Quasiparticle Band Structure in Monolayer MoS$_2$

Yi Lin[1,^], Yang-hao Chan[1,%], Woojoo Lee[2], Li-Syuan Lu[4,5], Zhenglu Li[1], Wen-Hao Chang[4,5], Chih-Kang Shih[2], Robert A. Kaindl[1,*], Steven G. Louie[1,3], Alessandra Lanzara[1,3,#]

[1] Materials Sciences Division, Lawrence Berkeley National Laboratory, Berkeley, California 94720, USA
[2] Department of Physics, The University of Texas at Austin, Austin, Texas 78712, USA
[3] Department of Physics, University of California, Berkeley, California 94720, USA
[4] Department of Electrophysics, National Yang Ming Chiao Tung University, Hsinchu 30010, Taiwan
[5] Research Center for Applied Sciences, Academia Sinica, Nankang, Taipei 11529, Taiwan

^ Corresponding author: yilin@lbl.gov
# Corresponding author: ALanzara@lbl.gov
% Current address: Institute of Atomic and Molecular Sciences, Academia Sinica, Taipei 106, Taiwan
* Current address: Department of Physics and CXFEL Laboratory, Arizona State University, Tempe, Arizona 85287, USA

Abstract

Optical excitation serves as a powerful approach to control the electronic structure of layered Van der Waals materials via many-body screening effects, induced by photoexcited free carriers, or via light-driven coherence, such as optical Stark and Bloch-Siegert effects. Although theoretical work has also pointed to an exotic mechanism of renormalizing band structure via excitonic correlations in bound electron-hole pairs (excitons), experimental observation of such exciton-driven band renormalization and the full extent of their implications is still lacking, largely due to the limitations of optical probes and the impact of screening effects. Here, by using extreme-ultraviolet time-resolved angle-resolved photoemission spectroscopy together with excitonic many-body theoretical calculations, we directly unmask the band renormalization effects driven by excitonic correlations in a monolayer semiconductor. We revealed a surprising bandgap opening, increased by 40 meV, and a simultaneous enhancement of band effective mass. Our findings unmask the novel exciton-driven mechanism towards the band engineering in photoexcited semiconducting materials, opening a new playground to manipulate the transient energy states in layered quantum materials via optical controls of excitonic many-body correlations.

Main

Excitons, excited bound electron-hole pairs, have been widely observed in layered two-dimensional materials and have attracted significant interests, given their critical roles in both fundamental science and applications. With engineered dimensionality, dielectric environment, and excitation density, excitons and its derivative many-body quasi-particles have played significant roles in the functionality of optoelectrical and energy-harvesting devices[1-3] and in the realization of exotic quantum phases involving Mott physics[4-7], charge density wave formations[8-10] and Bose-Einstein condensations[10-12]. However, due to the intricate many-body nature of the excitons, it remains an open question whether excitonic correlation renormalizes the single-particle band structure of the materials, as it is the case for other many-body interactions, such as electron-phonon[13,14], electron-electron[15] and electron-plasmon[16], as well as whether this could provide an inspiring new direction of engineering matters in correlated phases.

The theoretical framework for addressing excitonic correlations by using the T-matrix self-energy in screened-ladder approximation was proposed four decades ago[17] and has often been used in recent theoretical studies of photoexcited layered semiconductors[18-22], which suggest extraordinary renormalization of the quasiparticle bandgap and dispersions. However, this type of band renormalization has been experimentally obscured. To provide experimental evidence of the exciton-driven band renormalization, we need a probe that can directly map the material's band structure in both energy and momentum, while also being able to probe exciton formation and dynamics. So far, most of optical experimental tools used to studying excitons[23] do not access directly the material's electron self-energy $\Sigma(\mathbf{k}, \omega)$ in momentum space, and hence are unable to study, if any, such band renormalization effects. In contrast, angle-resolved photoemission spectroscopy (ARPES) has been an ideal tool to measure materials band structure. Only recently, time-resolved ARPES experiments have revealed that specific signatures of such exciton bound state may indeed be present in the single-particle spectral function as part of the spectra in photoemission of layered tungsten dichalcogenides[24-27], inspiring new opportunities for studying the exciton-driven band structure renormalization and engineering in materials.

Among the various material candidates, monolayer (ML) transition metal dichalcogenides (TMDCs) have been demonstrated as a versatile platform for realizing diverse excitonic and many-body phenomena via optical excitation[28-30], chemical doping[31,32], and electrical gating[33,34], which constitutes an promising platform for exploring the exciton-driven band normalization. In addition, the versatility to place them on different substrates (from insulating to conducting), together with the diverse choice of optical excitation density, provides the flexibility to operate in different regimes of interacting many-body electron-hole gas, which helps unmask the exciton-driven band renormalization from the strong screening effects of free carriers [35-46].

In this work, we apply extreme-ultraviolet time-resolved angle-resolved photoemission spectroscopy (XUV-trARPES) with high energy, momentum, and time resolution and in the low pump fluence regime, to study the ultrafast electronic structure renormalization in monolayer $MoS_2$ on a substrate of conducting highly oriented pyrolytic graphite (HOPG) at 300K. With a high-repetition-rate XUV-trARPES system[47,48], we discover exciton-driven ultrafast bandgap dynamics and band renormalization at the K valley of the ML $MoS_2$. Combined with theoretical calculations, we provide evidence on the unique role that excitonic correlations play in inducing the exotic band renormalizations.

Figure 1(a) illustrates the schematic of our experiment. A highly oriented monolayer $MoS_2$/HOPG sample (see low-energy diffraction pattern in Fig. 1(b)) is excited by a visible 2.2 eV femtosecond pulse (pump, duration <130 fs) to create a transient excited state, and then probed by a high- harmonic-generation (HHG)-derived pulse (probe) at 22.3 eV via angle-resolved photoemission spectroscopy (ARPES) as a function of time. Data were taken along the Γ-K direction (see Brillouin zone sketch in Fig. 1(c)). More details on the experimental settings and sample preparation can be found in Sections 1 and 2 in the Supplementary Material (SM). Figures 1(d) and 1(e) present the band structure for negative (t = -500±100 fs) and positive time delay (t = 300±100 fs). Throughout the paper, negative delay times corresponds to equilibrium, before the pump pulse excites the sample. The most obvious change following the optical excitation is the appearance of the conduction band minima at K. To highlight finer changes, in Fig. 1(f) and 1(g), we compare the equilibrium and excited state energy distribution curves (EDCs) at the center of the K valley k=K (-1.29 Å$^{-1}$), corresponding to the momentum location of valence band maximum (VBM) and conduction band minimum (CBM), and at a wavevector location away from K where the conduction band feature in the excited state is barely visible (we call the edge of the K valley) k=$k_{valley\ edge}$ (-1.14Å$^{-1}$). The equilibrium spectra at the K valley center (black curve in Fig. 1(f)) shows a splitting of the valence band peak around -1.8 eV (relative to the Fermi energy), denoted by the Greek letter β, due to the well-known spin-orbit coupling [46]. Following the pump excitation, the conduction band gets populated and a new peak (α), corresponding to CBM, appears around 0.2 eV (orange curve). This is accompanied by a downshift of the valence band maximum, followed by a depletion of its spectral weight, which makes it harder to resolve the spin-orbit splitting.

The ability to resolve both the CBM (α) and VBM (β) with significantly advanced data quality, compared to prior XUV-trARPES studies of layered MoS$_2$ [37,38,49-53], makes our experiment powerful to study the dynamics of the gap in response to excitation in a quantitative way. Fig. 2(a) and 2(e) show the color maps of the spectral intensity as a function of delay time and binding energy at the K point and k=$k_{valley\ edge}$, taken along the vertical orange and green lines in Fig. 1(e), respectively. Following the pump pulse (t>0), a feature above Fermi energy appears, which from our interpretation below corresponds to the CBM for k=K spectra (in Fig. 2(a)) and the CB edge for the k= $k_{valley\ edge}$ spectra (in Fig. 2(e)). This feature persists for picoseconds after the pump excitation. The excitation-induced energy position changes of both the conduction and valence bands can be extracted by fitting the EDC spectra with single or double peak functions (see methods of the fittings in the SM) and are shown as a function of delay time in panels b-c and f-g for k=K and k= $k_{valley\ edge}$, respectively. Immediately after excitation, both the higher branch of spin-split valence band and conduction band show a clear downshift in energy at the two momenta positions. The dynamics of the relative shift of VB and CB at these two **k** points can be directly extracted from these data. At the K point, the relative difference is nothing else than the direct band gap ($\Delta E_K$ = Gap) as shown in Fig. 2(d). Surprisingly, the data reveals an increase of the bandgap, up to 40 meV (from 2.02 eV to around 2.06 eV) following the optical excitation, and then the increased gap starts to recover with a decay rate τ = 5.0±1.8 ps. In contrast, at the edge of the discernable valley (k=$k_{valley\ edge}$), the energy separation between the CB and top VB ($\Delta E_{valley\ edge}$) decreases by ~100 meV. The opposite direction between $\Delta E_K$ and $\Delta E_{valley\ edge}$ suggests that non-monotonic **k**-dependent energy renormalization effects might be at play, as illustrated by the sketch in Fig. 2(i). It is important to note that our observed bandgap increase cannot be attributed to the known optical Stark or Bloch-Siegert effects [54,55], which lead to the increase of the optical gap (instead of bandgap) due to the optically driven coherence, requiring the persistent presence of a light field (instead of excitonic many-body correlations).

To gain further insights on the **k**-dependent band renormalization, in Fig. 3, we provide a quantitative analysis of the dispersion and effective mass of the conduction and valence band after photoexcitation. Figure 3(a-c) and 3(d-f) show the color maps of the spectral intensity, as a function of energy and momentum, at different characteristic delay times. In Fig. 3(a), at t=0 fs (defined by the center of the pump-probe cross-correlation), the pump excitation slightly populates the CB. At later delay time (panels b-c), the conduction band is gradually depopulated, after an initial increase of the photoemission intensity at the CBM due to the scattering of electrons from other k points. Fig. 3(g) and 3(h) report the dispersions of the CB and the upper branch of split VB, respectively, obtained by fitting the EDC spectra. The data show a clear flattening of the dispersion at later delay time (compare blue dashed curve (0 fs) with red curve (1.5ps)), leading to an increase of the CB effective mass by a factor of 3, from 0.71 $m_e$ to 2.45 $m_e$ ($m_e$ = electron mass in vacuum), as shown in Fig. 3(i). Such change occurs already within the first 1.5ps after excitation, corresponding to the time window of the fast-declining edge in Fig. 2(c). A similar, although smaller, flattening of the dispersion (panel h, compare grey and red curves) and hence enhancement of the effective mass (panel j), are also observed for the valence band. These apparently smaller enhancements might be affected by the decrease of the VB spectral weight in the transient state, making it hard to resolve the spin splitting of the two bands.

The results shown so far, point to the presence of strong photoexcitation-induced band renormalization, manifested by an increase of both the bandgap and effective mass. As discussed in detail below, we attribute these observations to the exciton-driven band renormalization, as schematically illustrated in Fig. 4(a). Clearly these exciton-driven effects are in striking contrast to the ones driven by the unbound photoexcited free carriers in Fig. 4(b), such as the large bandgap decrease that is attributed to the strong screening effects due to the high density of excited e-h gas as free carriers. Such bandgap decrease has been explained by theoretical tools such as GW methods[56,57] and has been observed in (tr)ARPES studies of monolayer semiconductors by injecting high-density electron dopants[32] or high-fluence photoexcitation[37,49].

However, it is critical to note that the free-carrier-driven mechanism alone does not fully account for our observations. Indeed, the unbound free carriers are expected to only drive a monotonic decrease of the bandgap for increasing carrier densities, and therefore cannot explain our observation of the bandgap increase. Moreover, the excitation fluences in our experiment are in the range of a few tens of μJ/cm$^2$ (see methods section), which is about 10-100 times lower than the fluences adopted in previous XUV-trARPES work on MoS$_2$ [37,38,49-53] (hundreds μJ/cm$^2$ to mJ/cm$^2$), The resulting excitation densities in our experiment are expected to be beneath the Mott threshold which is believed to be mid to high $10^{12}$ cm$^{-2}$ for ML MoS$_2$ [49].

To gain a full picture of what might drive the non-trivial renormalization effects here reported, it is necessary to consider the bound electron-hole pairs with excitonic correlations. To this end, we perform many-body theoretical calculations going beyond the free-carrier-driven picture. With the assumption that bound excitons with excitonic correlations are present in the excited electron-hole gas below Mott threshold, we calculate the interacting single-particle spectral functions of the ML MoS$_2$ at various excitation densities using a self-energy formalism, where the GW diagram and the T-matrix ladder diagrams are integrated to include the exciton-driven effects. The details of the self-energy construction are available in the Section 3.2-3.4 of SM. We further employ the quasi-equilibrium theory which assumes that the dynamics of the photoexcited electrons and holes changes steadily after the optical excitation (see details in the Section 3.1 of SM). Within this picture, we simulate the band renormalization as a function of densities of photoexcited electrons and holes.

In Fig. 4(c), we present the calculated quasiparticle bandgap of ML MoS$_2$ as a function of the excited e-h density. The inclusion of the excitonic correlations in the simulation leads to an appealing non-monotonic behavior of the gap-density curve. When a small finite number of electrons and holes are excited and available to form excitons, we observe a sudden increase of the calculated band gap, up to several tens of millielectronvolts, followed by a decrease, as the carrier density further increases. The dome-like increased-gap area at low density regime is the result of excitonic correlations and is responsible for the observation of the gap increase in Fig. 2(d).

To unmask experimentally the exciton-driven gap increase, we also need to consider the presence of the substrate effects. Environmental screening effects modify the effective dielectric constant in the sample via substrates or by encapsulating the 2D layer with other materials [58-60]. In the case of a conducting substrate, such as the HOPG utilized in this study, the charge in the substrate screens the monolayer semiconductor and is known to drive a reduction of the quasiparticle bandgap up to ~200 meV from its free-standing value [58]. The bandgap reduction from that of a free-standing sample (theoretical bandgap ~2.4 eV) is indeed observed in our experiment, where the direct gap size with maximum populated CBM is 2.10 ± 0.02 eV at 80K, in line with that in previous STS work on a similar substrate at 77K [61,62]. The fact that the gap size is comparable in both experiments, i.e., when measured without photoexcitation (as it is the case in equilibrium STS experiments) or with photoexcitation (as in this trARPES study), confirms the role of the conducting substrate in modulating the response of the bandgap to photoexcitation.

To quantify the substrate effects and how this is reflected in our findings, in Fig. 4(d), we present the calculated quasiparticle bandgap of ML $MoS_2$ on HOPG substrate as a function of the photoexcited e-h density. The gap-density curve presents a significantly gentler slope in the descending tail, compared to that of the freestanding case in Fig. 4(c), which results in a broader dome of increased gap toward higher excitation density. The gentler slope could be understood in a way that the bandgap has been environmentally screened by the conducting substrate, so it becomes less subjective to the additional screening-induced gap decrease due to the photoexcited free carriers. Based on the calculated results, we conclude that the presence of the HOPG substrate helps extend the increased-gap dome over a broader range of excitation densities, providing a more ideal situation to experimentally search for the exciton-driven bandgap increase.

Another consequence of the excitonic correlations in theory is the band renormalization in the proximity of the CBM and VBM as shown in Fig. 4(e) for two representative excitation densities. Besides the increase of the band gap discussed above, the calculations reveal a flattening of VB and CB following exciton formation (red curve) with respect to the equilibrium dispersion (black curves). This trend qualitatively agrees with our experimental observation of decreased effective mass, discussed in Fig. 3.

In conclusion, we have provided experimental and theoretical evidence of exciton-driven renormalization effects in the electronic band structure of the monolayer semiconductor, leading to a transient enhancement of the bandgap and effective mass. The combination of the appropriate regime of excitation strength, the usage of the conducting substrate and the supreme data quality have enabled us to unmask, for the first time, the exciton-driven band renormalization effects, previously obscured. The excitonic mechanism can be applied to understand the emerging non-monotonic bandgap behavior in wide range of layered and bulk excitonic semiconducting and insulating materials [62-64] and opens up new directions and pathways for optically-driven ultrafast band engineering.

Acknowledgement:

Y.L., A.L and R.A.K acknowledge the support from U.S. Department of Energy (DOE), Office of Science, Office of Basic Energy Sciences, Materials Sciences and Engineering Division under Contract No. DE-AC02-05-CH11231 (Ultrafast Materials Science program KC2203), which is the primary fund for the experimental work.  Y.H.C., Z.L. and S.G.L. acknowledge support from U.S. Department of Energy (DOE), Office of Science, Office of Basic Energy Sciences, Materials Sciences and Engineering Division under Contract No. DE-AC02-05-CH11231 (Theory of Materials Program KC2301), which is the primary fund for the theoretical work.  The computational part of this work used the following computing centers: Stampede2 at the Texas Advanced Computing Center (TACC) supported by National Science Foundation (NSF) under Grant No. ACI-1053575; Cori at National Energy Research Scientific Computing Center (NERSC) supported by the Office of Science of the DOE under Contract No. DE-AC02-05CH11231; Frontera at TACC supported by NSF under Grant No. OAC-1818253. Y.H.C. thanks C.S. Ong for helpful discussion. C.K.S. and W.J.L. acknowledge the support from the NSF Center for Dynamics and Control of Materials with grant No. DMR1720595, NSF-DMR 1808751, Welch Foundation F-1672 and US Airforce FA2386-18-1-4097. W.-H.C. acknowledges the supports from the Center for Emergent Functional Matter Science (CEFMS) of National Yang Ming Chiao Tung University and from the Ministry of Science and Technology (MOST) of Taiwan (107-2112-M-009-024-MY3 and 108-2119-M-009-011-MY3).

Author Contributions:

A.L. and Y.L. initiated and directed this research project. Y.L. and W.L. carried out the tr-ARPES and LEED measurements. Y.-H.C. performed theoretical calculations, with supervision from S.G.L.; Y.-H.C., Z.L. and S.G.L. analyzed the theoretical results. The sample was prepared by L.-S.L. and W.L. for the measurement, with supervision from C.-K.S and W.-H.C., Y.L. built the OPA dedicated to this work and upgraded the pump-probe scheme from an XUV-trARPES setup developed by R.A.K., Data analysis were performed by Y.L. with help from A.L. The paper was written by Y.L. and A.L. with critical input from all other authors.

Figures:

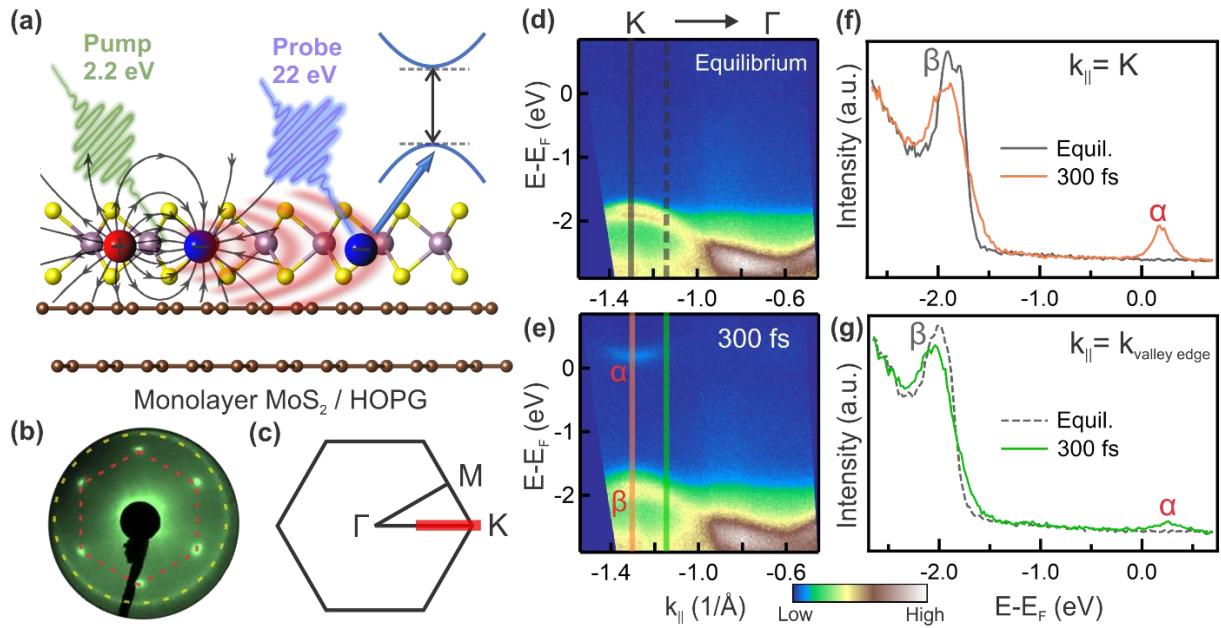

Figure 1. (a) A schematic of HHG-trARPES experimental settings. (b) LEED characterization of ML MoS$_2$/HOPG surface. Red (yellow) dashed lines denote the diffraction patterns from ML MoS2 (HOPG). (c) Brillouin zone and high symmetry points of ML MoS$_2$. Red solid line highlights the region where band structures are experimentally measured. Band structure measured for (d) equilibrium (t = -500±100 fs) and for (e) excited (t = 300±100 fs). Vertical lines in (d) and (e) denote the cuts at fixed momenta where energy distribution curves (EDCs) are extracted. (f) EDCs at K point (k=1.29 Å$^{-1}$) along the solid black line in (d) and the orange line in (e). (f) EDCs at k=1.14 Å$^{-1}$ along the dashed black line in (d) and the green line in (e). Greek letters in (e)-(g) highlight the major spectral features discussed in the manuscript.

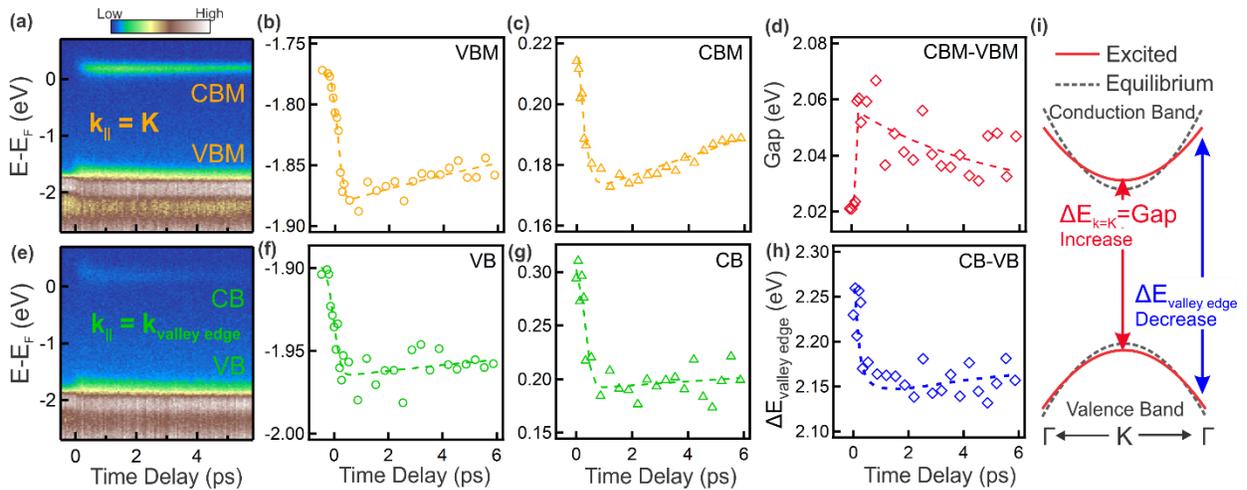

Figure 2. Color map of the spectral intensity as a function of delay times and binding energy at (a) K point (k=-1.29 ± 0.01Å$^{-1}$) and at (e) K valley edge (k=-1.14 ± 0.01 Å$^{-1}$). Energy-fitted band position in time for (b) VBM and (c) CBM at K point, and for (f) top VB and (g) CB at k=-1.14 Å$^{-1}$. (d) Bandgap dynamics at K. (h) Band energy distance dynamics at K valley edge (k=-1.14 Å$^{-1)}$. Dashed lines in (b)-(h) are guide for the eye. (i) A schematic of non-trivial band structure renormalization manifesting the non-monotonic **k**-dependent band energy distance dynamics in (d) and (h).

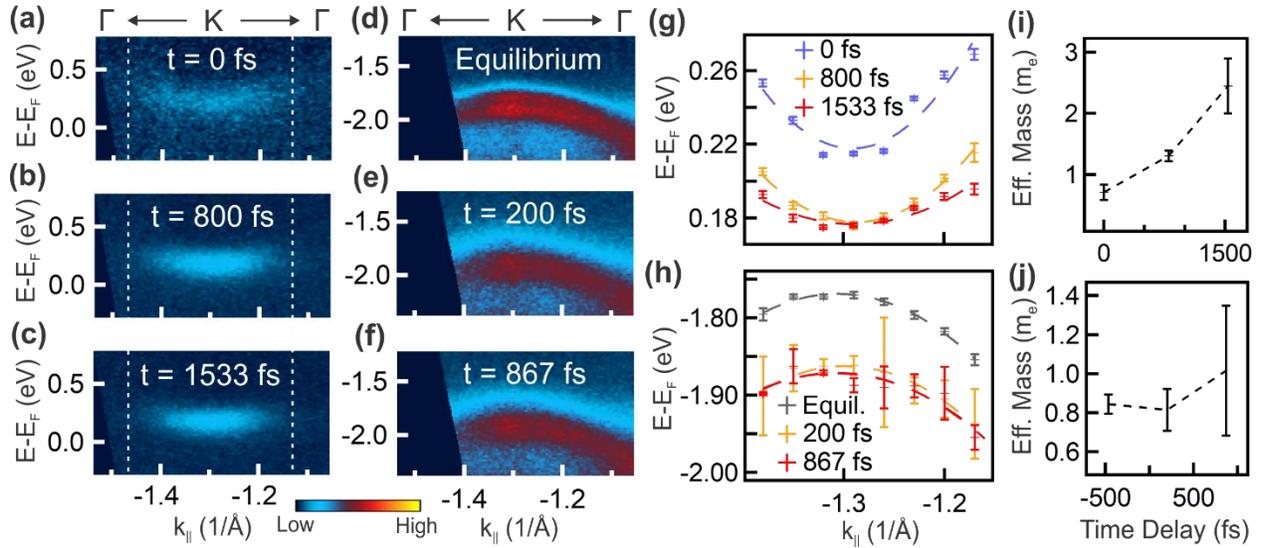

Figure 3. Dispersion near CBM for (a) t=0 fs, (b) t=800 fs, and (c) t=1533 fs. Dispersion of two bands near VBM for (d) equilibrium (t=-467 fs), (e) t=200 fs, and (f) t=867 fs. (g) Parabolic curve fits for CB dispersions in (a)-(c). (h) Parabolic curve fits for upper VB dispersions in (d)-(f). (i) Electron effective mass for the CB dispersions extracted from (g). (j) Hole effective mass for the upper VB dispersions extracted from (h).

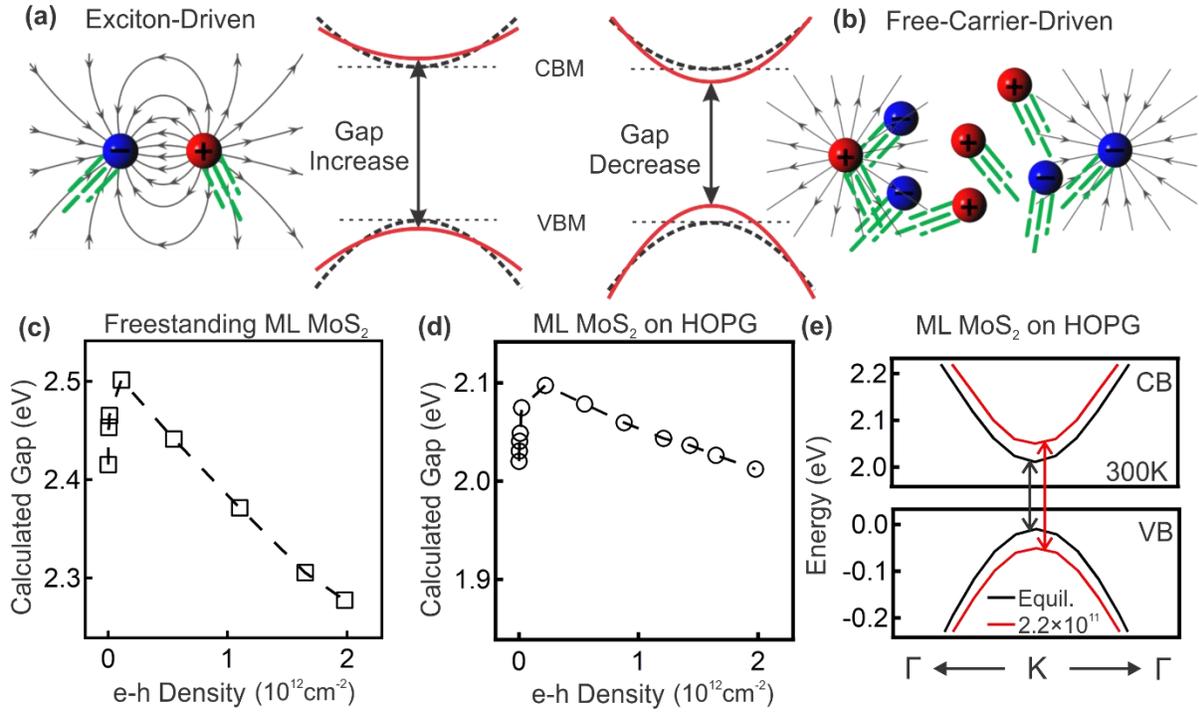

Figure 4. A schematic of (a) exciton-driven band gap increase and (b) free-carrier-driven band gap decrease. Calculated gap as a function of e-h density for (c) freestanding ML $MoS_2$ and (d) ML $MoS_2$ on HOPG. (e) Calculated CB and VB dispersions at two representative e-h density for ML $MoS_2$ on HOPG in 300K. Black and red arrows denote the bandgap at equilibrium and the excited phases.